\documentclass[superscript address,twocolumn,pra]{revtex4-2}
\usepackage{graphics}
\usepackage{graphicx}
\usepackage{bm}
\usepackage{amsmath}
\usepackage{mathtools}
\usepackage{color}
\usepackage{amssymb}

\begin{document}

\title{Photon emission without quantum jumps}
\author{Thomas Hartwell}
\affiliation{School of Chemical and Process Engineering, University of Leeds, Leeds, LS2 9JT, United Kingdom}
\affiliation{School of Physics and Astronomy, University of Leeds, Leeds, LS2 9JT, United Kingdom}
\author{Daniel Hodgson}
\affiliation{School of Physics and Astronomy, University of Leeds, Leeds, LS2 9JT, United Kingdom}
\author{Huda Alshemmari}
\affiliation{School of Physics and Astronomy, University of Leeds, Leeds, LS2 9JT, United Kingdom}
\affiliation{School of Chemical and Process Engineering, University of Leeds, Leeds, LS2 9JT, United Kingdom}
\author{Gin Jose}
\affiliation{School of Chemical and Process Engineering, University of Leeds, Leeds, LS2 9JT, United Kingdom}
\author{Almut Beige\footnote{Email: a.beige$@$leeds.ac.uk}}
\affiliation{School of Physics and Astronomy, University of Leeds, Leeds, LS2 9JT, United Kingdom}

\date{\today}

\begin{abstract}
When modelling photon emission, we often assume that the emitter experiences a random quantum jump. When a quantum jump occurs, the emitter transitions suddenly into a lower energy level, while spontaneously generating a single photon. However, this point of view is misleading when modelling quantum optical systems which rely on far-field interference effects for applications like distributed quantum computing and non-invasive photonic quantum sensing. In this paper, we highlight that the dynamics of an emitter in the free radiation field can be described by simply solving a Schr\"odinger equation based on a locally-acting Hamiltonian without invoking the notion of quantum jumps. Our approach is nevertheless consistent with quantum optical master equations.
\end{abstract}

\maketitle

\section{Introduction}

The common view of an individual, initially excited emitter is that it is capable of spontaneously releasing its energy while generating a single photon. This process seems inherently probabilistic and is often referred to as a {\em quantum jump} \cite{Zoller,Moelmer,Hegerfeldt,Carmichael}. Since it was initially incredibly difficult to observe an individual jump in the dynamics of an emitter, carefully designed ion trap experiments have instead been used to demonstrate the existence of so-called macroscopic quantum jumps \cite{jumps4,jumps5,jumps3}. These occur in the fluorescence of an emitter with a strongly-driven, rapidly decaying excited state and a weakly-driven, metastable state and manifest themselves as a random telegraph signal of long light and dark periods \cite{jumps,jumps2,jumps6}. Once the emitter transitions into the metastable state, it cannot emit light and might remain dark for a significant amount of time. In contrast to this, the continuous emission of light indicates that the metastable energy level is not populated. In this case, the metastable state is known to be unpopulated and might remain so for a very long time due to the quantum Zeno effect \cite{Beige}. The experimental observation of these macroscopic light and dark periods in the 1980s, despite some initial criticism of this interpretation \cite{Ballentine2,Ballentine3}, eventually manifested the belief that spontaneous photon emission and quantum jumps are two closely related phenomena. 
 
\begin{figure}[t]
 \includegraphics[width=0.99 \linewidth]{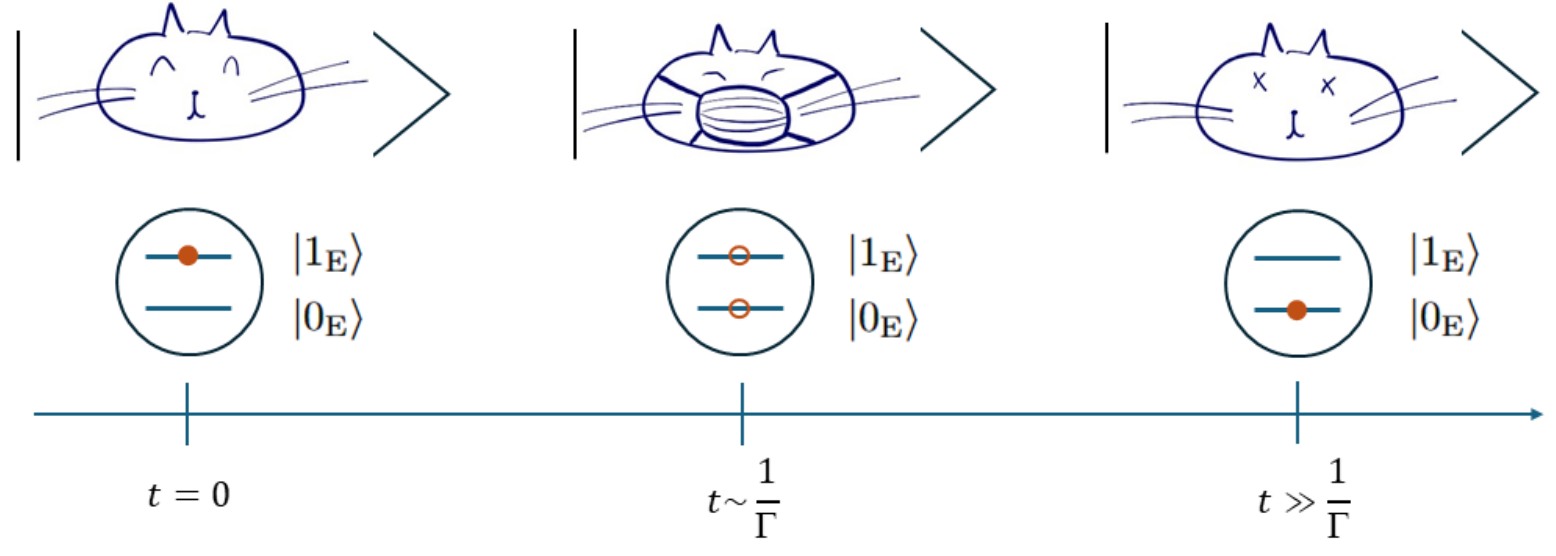}
     \caption{Schematic illustration of the generation of a single photon by an initially excited emitter with two internal electronic states. Suppose the excited state $|1_{\rm E} \rangle$ of the emitter corresponds to an {\em alive} cat, while $|0_{\rm E} \rangle$ denotes its ground state and corresponds to a {\em dead} cat. Utilising an analogy with Schr\"odinger's cat, the ability to treat the emitter and its surrounding radiation field as a closed quantum system which can be analysed with the help of a Schr\"odinger equation implies that the cat is in general both dead and alive and can transition continuously from being alive to being dead. This is in contrast to the common view which suggests that the cat is always either alive or dead.}
    \label{fig1}
\end{figure}
 
There are, however, other experiments that contradict this point of view and suggest that a quantum jump only occurs when a photon arrives at a detector. An example is the famous two-atom double-slit experiment \cite{Scully,Eichmann} which demonstrates that the light coming from atomic emitters is capable of interfering in the far-field, i.e.~long after it has been created. Such far-field interference is only possible if the collapse of the state of a quantum system only occurs when a measurement is performed and the quantum state needs updating according to the information that has been gained \cite{Schoen,Schoen2}. On a coarse grained time scale, the individual trajectories of an emitter with macroscopic quantum jumps can be seen as a series of individual quantum jumps due to the presence of an observer who performs actual fluorescence measurements \cite{Beige}. In general more care is needed when modelling the generation of individual photons in order to incorporate far-field interference effects with the ability to generate atomic long-range interactions \cite{Dawson2021}.

In quantum optics, we usually describe the dynamics of quantum systems with spontaneous photon emission by so-called master equations \cite{agarwal1970,petro}. These describe the time evolution of the density matrix of the emitter and can be used to predict the expectation values of measurements. Quantum optical master equations are, however, not very intuitive and their derivation usually requires several ad hoc assumptions and  approximations, such as the rotating wave, the Born and the dipole approximations \cite{stokes2012}. In addition, master equations do not tell us how to unravel the dynamics of the atomic density matrix into the quantum trajectories seen in experiments with individual quantum systems \cite{Moelmer,Hegerfeldt,Carmichael}. In this paper, we therefore have a fresh look at a single emitter inside a free radiation field. Inspired by Ref.~\cite{Axel} and as illustrated in Fig.~\ref{fig1}, we liken the generation of a single photon in the following to a Schr\"odinger's cat \cite{Erwin} which transitions continuously from being {\em alive} (emitter excited and no photon present in the surrounding free radiation field) to being {\em ill} (emitter and field both excited) until it eventually becomes {\em dead} (emitter in the ground state and excitation present in the field) instead of jumping spontaneously from one quantum state into  another.

Experiments have shown that the internal dynamics of a single-photon emitter with only two internal energy eigenstates is relatively simple: the excited state population simply decreases exponentially in time. This suggests that it should be possible to model photon emission in a much more straightforward way than using master equations \cite{agarwal1970,petro,stokes2012}. As we shall see below, a single-photon emitter is essentially a {\em closed} quantum system with a Hamiltonian of the form
\begin{eqnarray} \label{E1}
H &=& H_{\rm E} + H_{\rm F} + H_{\rm int} 
\end{eqnarray}
where $H_{\rm E}$ and $H_{\rm F}$ denote the Hamiltonian of the emitter and the surrounding free radiation field, respectively, and $H_{\rm int}$ captures their interaction. By simply solving the corresponding Schr\"odinger equation, we find that single-photon emitters essentially resemble classical antennae \cite{Pohl} connected to finite-sized batteries. The energy of the circuit is continuously released into the free radiation field with an intensity that is proportional to the energy left in the battery. The re-absorption of released photons does not occur, since, once emitted, local excitations of the electromagnetic field move away from their source at the speed of light.

As we highlight below, there are no quantum jumps unless an actual measurement is performed either on the emitter or on the free radiation field. Hence, until they are eventually detected, emitted photons can interfere \cite{Schoen2}. This observation has potential applications in quantum technology that range from distributed quantum computing \cite{RUS,Kok} to non-invasive photonic quantum sensing \cite{Dawson2021}. Indeed the observation that the generation of a single photon is a {\em coherent} process is not new. Fedorov {\em et al.} \cite{Eberly} and more recently Longhi \cite{Longhi} showed that Hamiltonians of the same general form as $H$ in Eq.~(\ref{E1}) are consistent with an exponential decrease of the excited-state population of a two-level emitter. The main difference between our paper and previous papers, like Refs.~\cite{Eberly,Longhi}, is that we describe the quantised electromagnetic field in the following in position and not in momentum space.

In general, it is assumed that the quantum state of an individual photon, which is the elementary particle of the electromagnetic field, is a superposition of monochromatic single field-excitation states. A complementary approach is to decompose the wave packet of a single photon into {\em local} excitations which have a unique position at any given instance in time \cite{photons,Jake2021,Gabriel2025}. In the past, it has been widely believed that a local description of the quantised electromagnetic field is not possible, since local electric field observables do not commute with each other even when referring to different positions. However, previous no-go theorems \cite{Sipe} can be overcome by distinguishing between electromagnetic field observables and the local carriers of light. The annihilation and creation operators of these local carriers of light are bosonic and commute when referring to different positions \cite{photons}. As we shall see below, considering local photons allows us to avoid many approximations and ad hoc assumptions of standard approaches to the modelling of quantum optical systems with photon emission \cite{Moelmer,Hegerfeldt,Carmichael,agarwal1970,petro,stokes2012}.

In addition to locality, the construction of the interaction Hamiltonian $H_{\rm int}$ takes the experimental observation into account that the transition of an emitter from one energy level into another results in the generation of exactly {\em one} photon \cite{Rempe}. Hence the interaction Hamiltonian of a point-like emitter with two internal states must be of the form
\begin{eqnarray}  \label{InteractionHamiltonian}
H_{\rm int} &=& \hbar g \, a^\dagger (0) \sigma^- + {\rm H.c.}
\end{eqnarray}
Here $g$ represents a coupling constant, $\sigma^-$ denotes the atomic lowering operator, and $a^\dagger (0)$ is the creation operator of a local excitation of the electromagnetic field at the position of the emitter. In other words, we assume that the emitter does not couple in a certain way to the observables of the quantised electromagnetic field but generates the local building blocks of a single photon, so-called {\em blips} (bosons localised in position). Imposing locality and considering the above interaction Hamiltonian, we are able to avoid the usual dipole approximation \cite{multi,multi2,multi3}. The above interaction Hamiltonian ensures causality in Fermi's famous two-atom problem \cite{Heg2,Milonni}. Moreover, as long as local blip excitations are only generated sufficiently slowly, the electromagnetic field can surround blips in a similar non-local way as a gravitational field surrounds a massive object without violating non-locality, i.e.~without spreading faster than allowed by the speed of light.

Eq.~(\ref{InteractionHamiltonian}) also takes into account that our quantum description of emitter and field is consistent with the second law of thermodynamics which forbids the flow of energy from a ``colder" to a ``hotter" subsystem. As shown in Refs.~\cite{stokes2017,RevMod}, quantum subsystems need to be defined such that their interaction Hamiltonian commutes with the interaction-free Hamiltonian, 
\begin{eqnarray}  \label{InteractionHamiltonian2}
\left[ H_{\rm E} + H_{\rm F}, H_{\rm int} \right] &=& 0 \, .
\end{eqnarray}
Indeed there is an ambiguity regarding the identification of quantum subsystems, i.e.~what we call the emitter and what we call the surrounding quantised electromagnetic field. This applies, since any Hamiltonian $H'$ which relates to $H$ in Eq.~(\ref{InteractionHamiltonian}) via a unitary transformation $U$ such that $H'=U H U^\dagger$ is unitarily equivalent and has the same energy spectrum. However, introducing $U$ changes the interaction Hamiltonian $H_{\rm int}$ into $H_{\rm int}'=U H_{\rm int} U^\dagger$. For example, $H_{\rm int}'$ might contain a counter-rotating term of the form $a^\dagger (0) \sigma^+ $ which can result in the emission of a photon, even when the emitter and the free radiation field are initially both in their respective ground states. This does not seem to be the case in actual quantum optics experiments \cite{Kurcz}. Usually, this problem is avoided with the help of the so-called rotating wave approximation which removes all counter-rotating terms and subsequently leads to an interaction Hamiltonian similar to the one in Eq.~(\ref{InteractionHamiltonian}).

A further approximation that is usually required when analysing the dynamics of a point-like two-level system with photon emission is the so-called Wigner-Weisskopf approximation \cite{WW,Stenholm,Berman}. This approximation too is avoided here, since $H_{\rm int}$ in Eq.~(\ref{InteractionHamiltonian}) is different from the interaction Hamiltonian that quantum opticians usually consider when modelling photon emission. Due to locality, its coupling constant $g$ is {\rm not} frequency-dependent. Moreover, obtaining a local description of the quantised electromagnetic field requires a doubling of its Hilbert space. Using a physically-motivated approach to quantisation \cite{photons,Jake2021,Gabriel2025}, it is noticed that the configuration space of light must support monochromatic waves with positive {\em and} with negative frequencies in order to accommodate localised wave packets of any shape and with any possible direction of propagation \cite{Ornigotti}. This means, that if we were to perform our calculations in momentum and not in position space, we would arrive at frequency integrals whose limits extend automatically from minus to plus infinity and can be solved analytically in a straightforward way. In summary, our approach to the modelling of photon emission avoids several quantum optical standard approximations and ad hoc assumptions. Our only assumption in this paper is the identification of the emitter with a point-like two-level system.

The main purpose of this paper is to obtain a more intuitive picture of the emission process. To achieve this, we highlight that quantum optical systems with photon emission are essentially closed quantum systems and that their dynamics can be predicted analytically by solving a Schr\"odinger equation based on a locally-acting interaction Hamiltonian. Nevertheless, our calculations can be used to obtain a master equation for the dynamics of the density matrix of the emitter by tracing out the field. Here this is justified by only being interested in the properties of the emitter, while the field degrees of freedom are ignored. However, taking a closed system approach, we can also ask about the properties of the emitted light. As an example, we calculate in the following the spectrum of the emitted light by performing a Fourier analysis on the state of the photon after it has become disentangled from its source. In good agreement with experimental observations \cite{Mollow,Mollow2,Mollow3}, we obtain a Lorentzian spectrum.

In the absence of emitter-field interactions, the excited state of the emitter accumulates a phase factor ${\rm exp}(-{\rm i} \omega_0 t)$ due to its free evolution, where $\omega_0$ denotes the atomic transition frequency. When a blip is generated, this phase factor is transferred onto the corresponding term in the state vector of the emitter-field system and subsequently remains the same. This means, the blips carry the phase factors of the emitter at the time of their creation. Hence the emitted photon seems to ``oscillate" the transition frequency of the emitter. Hence the spectrum of the emitted light is centred around $\omega_0$. Its broadening is due to an exponentially decreasing amplitude of the photonic wave function. 

Most importantly, the methodology presented in this paper opens the path to modelling photon emission in more complex scenarios, like atomic emitters in dielectric and plasmonic sub-wavelength cavities and emitters in the presence of two-sided partially transparent mirror interfaces \cite{Dawson2021,Abeer2}. Especially delays and far field interference effects can be taken into account more directly without relying, at least in principle, on classical response functions and other semi-classical approximations. In addition, our manuscript suggests new methods to preserve the state of emitters without the need for quantum feedback control. If the quantum state of the emitter is known at all times, is becomes in principle possible to apply laser driving to correct for any unwanted changes \cite{Ding}. In addition, our approach can be used to describe experiments which control the shape of photonic wave packets \cite{Axel2}.

This paper is organised as follows. The Results section shows that the excited-state population of the emitter decreases exponentially while it transfers its energy coherently into the surrounding field. Moreover, we find that the spectrum of the emitted light has a Lorentzian structure. Afterwards, we discuss the relation between our approach to modelling photon emission without quantum jumps and standard quantum optics models, like the quantum jump approach \cite{Moelmer,Hegerfeldt,Carmichael} and master equations \cite{agarwal1970,petro,stokes2012}, and summarise our findings. Finally, in the Methods section, we quantise the electromagnetic field originating from a single point-like source. 

\section{Results} \label{sec3}

\subsection{Dynamics of emitter and field} 

To analyse the dynamics associated with the Hamiltonian $H$ in Eq.~(\ref{E1}), we introduce the single-excitation states $|r_{\rm F} \rangle$ of the quantised electromagnetic field with $|r_{\rm F} \rangle = a^\dagger(r) \, |0_{\rm F} \rangle$. Here $|0_{\rm F} \rangle$ denotes the vacuum state and $a^\dagger(r)$ is a bosonic creation operator of field excitations which originated from a point-like source and radially travelled a fixed distance $r$ away from it, as described in Methods. In addition, we write the time evolution operator $U(t,0)$ of the emitter and the surrounding free radiation field as a Dyson series expansion (cf.~Eq.~(\ref{FinalDyson}) in Methods). Suppose the initial state of emitter and field is of the general form $|\psi(0) \rangle = \alpha \, |0_{\rm F},0_{\rm E} \rangle + \beta \, |0_{\rm F},1_{\rm E} \rangle$. Then we find that their state equals
\begin{eqnarray} \label{E24}
    |\psi(t) \rangle &=& \beta \, c_0(t) \, |0_{\rm F}, 1_{\rm E} \rangle + \beta \int_{-\infty}^{\infty} {\rm d}r \,c_r(t) \, |r_{\rm F}, 0_{\rm E} \rangle \notag \\
    && +  \alpha \, |0_{\rm F},0_{\rm E} \rangle  
\end{eqnarray} 
at all later times $t$. As Eqs.~(\ref{evenExponential}) and (\ref{E31}) in the Methods section show, the complex coefficients $c_0(t)$ and $c_r(t)$ in this equation equal
\begin{eqnarray}\label{oma}
    c_r(t) &=& - {\rm i} \, (\Gamma /c)^{1/2} \, {\rm e}^{({1 \over 2}\Gamma + {\rm i} \omega_0)(r/c-t)} \, , \notag \\
    c_0 (t) &=& {\rm e}^{- ({1 \over 2}\Gamma + {\rm i} \omega_0) t} 
\end{eqnarray}
for $0 \le r \le ct$ and with $\omega_0$ denoting the transition frequency of the emitter and with the spontaneous decay rate $\Gamma$ defined such that $\Gamma = g^2/c$. For $r<0$ or $t<0$, we have $c_r(t) = 0$ and $c_0(t)=1$. Moreover, $c_r(t) = 0$ when $r > ct$. The state vector $|\psi(t) \rangle$ denotes the pure state of the emitter and the surrounding free radiation field at any time $t>0$ under the condition that {\em no} measurement took place in $(0,t)$ which revealed any information about the emitter or the field. 

\begin{figure}[t]
   \includegraphics[width=1 \linewidth]{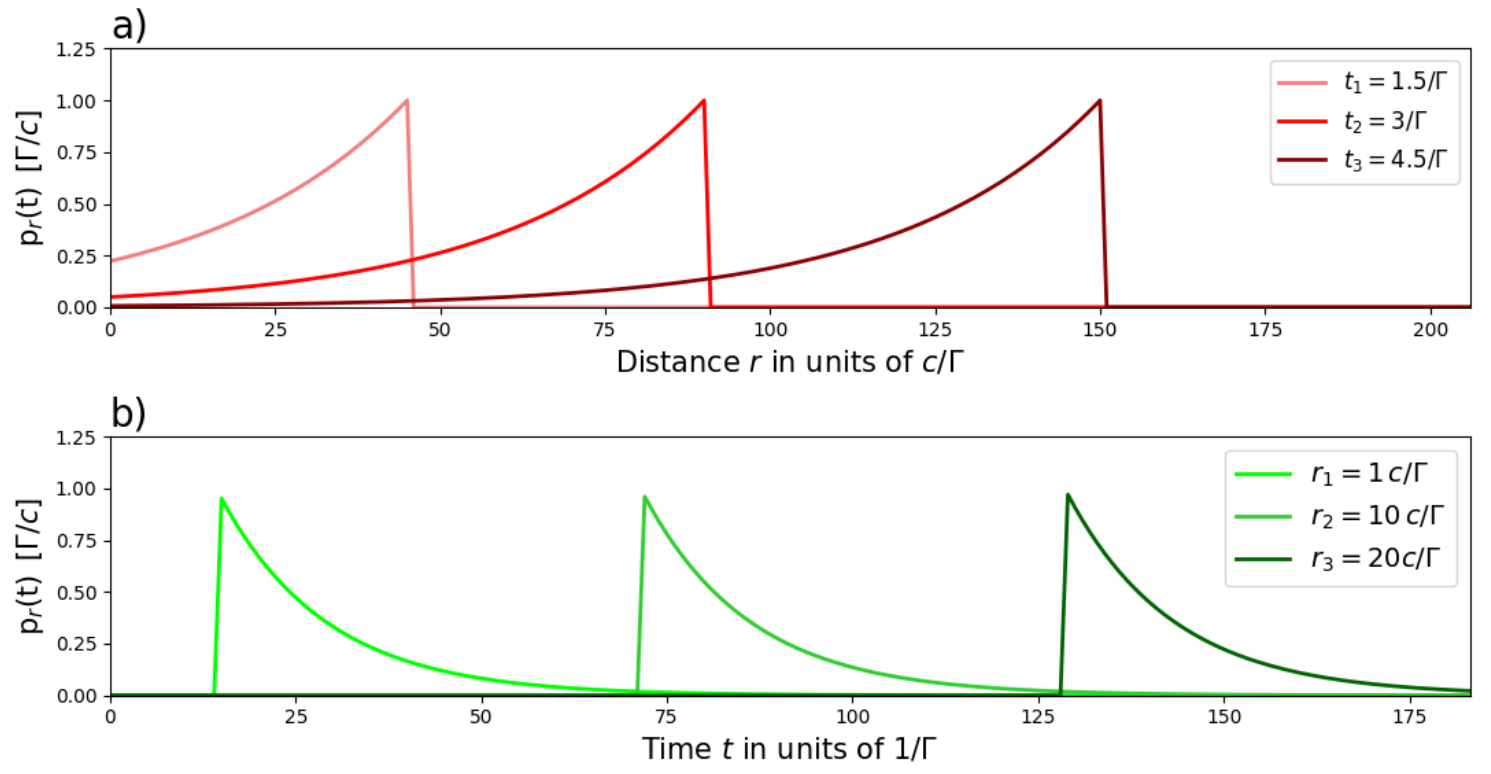}
    \caption{(a) Probability density $p_r(t)$ in Eq.~(\ref{oma2}) to detect a photon at time $t$ a distance $r$ away from an initially excited emitter ($|\beta|^2 = 1$) as a function of $r$ for three different times $t_1<t_2<t_3$. The figure shows that the generated photonic wave packet has an exponentially increasing amplitude and travels at the speed of light, $c$, away from the emitter. (b) The same probability density $p_r(t)$ as a function of the time $t$ for three different distances $r_1<r_2<r_3$. An observer placed at $r$ sees the wave packet arriving after some time $r/c$; afterwards its amplitude decreases exponentially in time.}
    \label{fig3}
\end{figure}

If a measurement is performed at any time $t$, then the state of emitter and field needs updating according to the information that is gained in the process. For example, the probability density of finding the photon emitter still in its excited state upon measurement at a given time $t$ is given by $p_0(t) = |c_0(t)|^2$. As we can see from Eq.~(\ref{oma}), 
\begin{eqnarray} \label{probCalc}
    p_{0}(t) &=& |\alpha|^2 + |\beta|^2 \, {\rm e}^{- \Gamma t} 
\end{eqnarray}
and its second term decreases exponentially and tends to zero as $t$ becomes much larger than $1/\Gamma$. This is as one would expect, since an initially excited emitter decays eventually. Moreover, suppose the emitter is {\em fully} surrounded by perfect photon detectors which are all a fixed distance $r$ away from the source. Having again a closer look at Eqs.~(\ref{E24}) and (\ref{oma}), we see that the probability density for any of the detectors to click at a given time $t$ equals $p_{r}(t) = |c_{r}(t)|^2$ with
\begin{eqnarray}\label{oma2}
    p_{r}(t) &=& (\Gamma/c) \,  |\beta|^2 \, {\rm e}^{\Gamma (r/c-t)} 
\end{eqnarray}
for $0 \le r \le ct$ and $ p_{r}(t) =0$ otherwise. The factor $1/c$ is needed here, since the $p_{r}(t)$ is a density per distance at a given time $t$ (cf.~Eq.~(\ref{E24})). When integrated over $r$, we find that $p_0(t) + \int_0^{ct} {\rm d}r \, p_r(t) = 1$. As illustrated in Fig.~\ref{fig3}, the emitter generates a single photon with an exponentially decreasing amplitude which moves outwards, away from its source, at constant speed. 

Despite not considering actual measurements, our approach is consistent with the quantum jump approach which introduces a conditional non-Hermitian Hamiltonian $H_{\rm cond}$ describing the dynamics of the emitter under the condition of no photon emission \cite{Moelmer,Hegerfeldt,Carmichael}. Usually, this Hamiltonian is obtained by considering environment-induced measurements on a coarse grained time scale $\Delta t$ which reveal information whether or not a photon has been generated. It was believed that $\Delta t$ needs to be chosen small enough to avoid the possible re-absorption of light by the emitter. Moreover, $\Delta t$ needs to be big enough to avoid the so-called quantum Zeno regime which would freeze the dynamics of the emitter. However, having a closer look at Eqs.~(\ref{E24}) and (\ref{oma}), we see that the conditional Hamiltonian $H_{\rm cond}$ of Refs.~\cite{Moelmer,Hegerfeldt,Carmichael},
\begin{eqnarray} \label{Hcond}
H_{\rm cond} &=& \hbar \left( \omega_0 - \textstyle{{\rm i} \over 2} \Gamma \right) |1_{\rm E} \rangle \langle 1_{\rm E}| \, ,
\end{eqnarray}
can be obtained without the assumption of environment-induced measurements which is in agreement with other authors, who argued that the dynamics of the emitter should not depend on the presence or absence of a distant observer \cite{Ballentine2,Ballentine3,Ballentine}. Indeed it does not matter whether the free radiation field is observed continuously, i.e.~on a coarse grained time scale $\Delta t$, or only once at a time $t$. The emitted light simply moves away from its source and therefore cannot be re-absorbed. Moreover, the predicted no-photon probability $p_0(t)$ in Eq.~(\ref{probCalc}) is the same in the presence and in the absence of environment-induced measurements. In addition, the quantum jump approach tells us that the state of the emitter is in its ground state if a photon is observed, which is also in agreement with the state vector given in Eq.~(\ref{E24}). However, there are also some differences. The quantum jump approach approximates the time $t-r/c$ by $t$ (cf.~e.g.~Eq.~(\ref{oma})), thereby neglecting the small amount of time it takes a photon to travel the distance $r$ from the source to the detector \cite{Hegerfeldt}. Notice also that our approach yields different predictions for quantum optical systems in which the emitted light can interfere before reaching the detector \cite{Dawson2021,Abeer2}. 

\subsection{The spectrum of the emitted light}

\begin{figure}[t]
  \includegraphics[width=1 \linewidth]{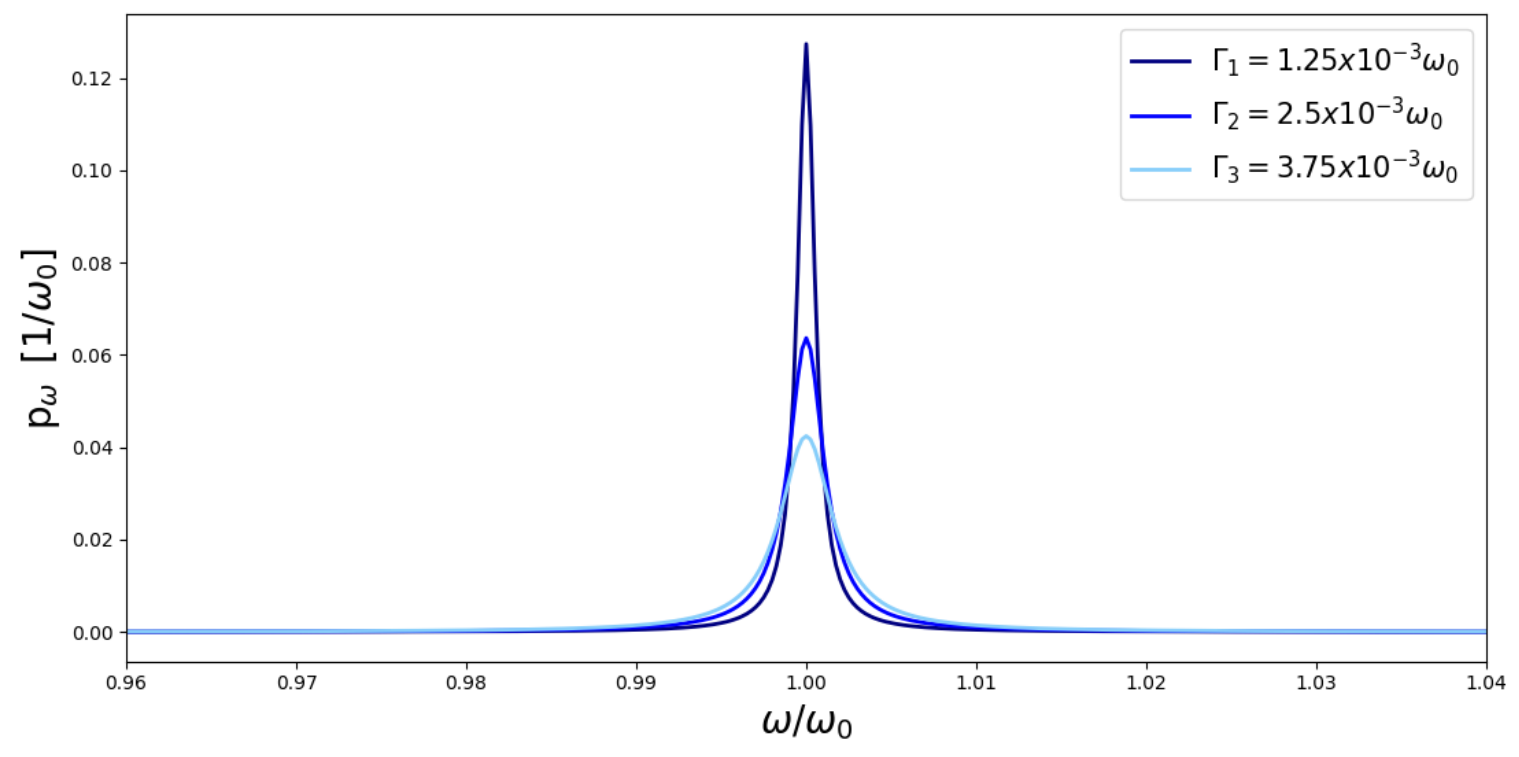}
    \caption{Probability density $p_{\omega}$ in Eq.~(\ref{oma22}) of the eventually emitted photon having the frequency $\omega $ for different decay rates $\Gamma_1 < \Gamma_2 < \Gamma_3$. Our calculations confirm in a relatively straightforward way that the spectrum of the emitted light is Lorentzian in agreement with experiments \cite{Mollow,Mollow2,Mollow3}.}
    \label{fig4}
\end{figure}

To verify that our calculations are consistent with experimental observations, we now have a closer look at the spectrum of the emitted light \cite{Mollow,Mollow2,Mollow3}. From Eq.~(\ref{E24}), we see that the state of the emitter and the field at time $t$ can also be written as
\begin{eqnarray} \label{E24new}
   |\psi(t) \rangle &=& \beta \, c_0(t) \, |0_{\rm F},1_{\rm E} \rangle + \beta \int_{- \infty}^\infty {\rm d}k \, \widetilde c_k(t) \, |k_{\rm F} , 0_{\rm E} \rangle \notag \\
   && + \alpha \, |0_{\rm F},0_{\rm E} \rangle 
\end{eqnarray}
with the single-excitation monochromatic state $ |k_{\rm F} \rangle$ defined such that $ |k_{\rm F} \rangle = \widetilde a^\dagger (k) \, |0_{\rm F} \rangle$. The coefficients $\widetilde c_k(t)$ relate to the $c_r(t)$ coefficients via a Fourier transform. Taking this into account and using Eq.~(\ref{oma}) and Eq.~(\ref{FieldFT1}) in Methods, we find that 
\begin{eqnarray} \label{E24new2}
   \widetilde c_k(t) 
   &=& {{\rm i} (c \Gamma/ 2 \pi)^{1/2} \over \Gamma/2 + {\rm i} (\omega_0 - ck)} \left[ {\rm e}^{- ({1 \over 2}\Gamma + {\rm i} \omega_0) t} - {\rm e}^{-{\rm i} ckt} \right] . ~~
   \end{eqnarray}
The probability density that the emitted photon has the frequency $\omega = ck$ is given by $p_{\omega}(t) = |\widetilde c_k(t)|^2/c$ and is time dependent until all light has left the emitter and $t \gg 1/\Gamma$. Hence $p_{\omega} =  \lim_{t \to \infty} p_{\omega}(t)$ becomes
\begin{eqnarray}\label{oma22}
   p_{\omega} &=& {1 \over 2 \pi} \cdot {\Gamma \over (\Gamma/2)^2 + (\omega_0 -\omega)^2} \, .
\end{eqnarray}
This shows that the spectrum of the emitted light is indeed Lorentzian \cite{Mollow,Mollow2,Mollow3}, as illustrated in Fig.~\ref{fig4}. The dominant frequency is  the transition frequency $\omega_0$ of the emitter, as one would expect. In addition, the standard deviation of this spectrum is proportional to the spontaneous decay rate $\Gamma$. However, notice that this result has been obtained here by simply solving the Schr\"odinger equation of emitter and field which is based on a locally-acting Hermitian interaction Hamiltonian without the need for approximations and ad hoc assumptions, like the assumption of complex eigenvalues or the need for continuous environment-induced measurements.

\subsection{Energy conservation}

The observable for the energy of the single-photon emitter is given by $H_{\rm E}$. Moreover, the energy observable $H_{\rm FE}$ of the free radiation field equals 
\begin{eqnarray} \label{eq:dynamical_hamnewextra}
H_{\rm FE} &=& \int_{-\infty}^{\infty}\text{d}k \,  \hbar c |k| \, \widetilde a(k)^\dagger \widetilde a(k) \, .
\end{eqnarray}
This operator has the same eigenvectors as the field Hamiltonian $H_{\rm F}$ but only positive eigenvalues \cite{Gabriel2025}. However, for sufficiently large spontaneous decay rates $\Gamma$, the emission rate $p_{\omega}$ for negative frequencies $\omega$ becomes negligible and $H_{\rm FE} \equiv H_{\rm F}$. Since $H_{\rm E} + H_{\rm F}$ commutes with the Hamiltonian $H$ in Eq.~(\ref{E1}), the time evolution in Eq.~(\ref{E24}) conserves the free energy of emitter and field. For sufficiently large times $t$, the energy of the emitted photon therefore coincides with the initial energy $\hbar \omega_0$ of the excited state of the emitter.

\section{Discussion} \label{sec4}

Quantum opticians usually consider an emitter with spontaneous photon emission to be an {\em open} quantum system. While the state vectors of closed quantum systems evolve unitarily according to the Schr\"odinger equation, open quantum systems need to be described by density matrices and evolve according to master equations. These can be derived phenomenologically or using second order perturbation theory, involving a variety of approximations and assumptions \cite{agarwal1970,petro,stokes2012}. Over the last decades, master equations have been widely used in the analysis of devices with quantum technology applications and their predictions have been found to be in good agreement with experiments. However, they often do not align well with our physical intuition, especially when emitters are placed in structured environments and far-field interference effects need to be taken into account. The analysis of more complex quantum optical systems can become very convoluted \cite{Dawson2021}. 

Motivated by these observations, this paper takes a more direct approach and demonstrates that an emitter placed inside the free radiation field is essentially a {\em closed} quantum system which remains at all times in a pure state $|\psi (t) \rangle$ (cf.~Eqs.~(\ref{E24}) and (\ref{E24new}) and Refs.~\cite{Eberly,Longhi}). Our analysis highlights that the emitter constantly creates local excitations, so-called {\em blips} which stands for bosons localised in position \cite{photons,Jake2021,Gabriel2025}, in the free radiation field. These cannot be re-absorbed by the emitter since they move away from the source at the speed of light. Each blip caries the phase of the emitter at the time of its creation. Hence the real parts of the electric and magnetic field amplitudes of the emitted light oscillate at the transition frequency $\omega_0$ of the emitter. Our predictions are in good agreement with the predictions obtained using alternative methods. For example, we observe that the emitter loses its initial excitation in an exponential fashion at a constant rate $\Gamma$, as illustrated in Fig.~\ref{fig3}(a). The generation of a single photon is in general not spontaneous. 

If we are only interested in the dynamics of the emitter, then the field degrees of freedom of the state $|\psi (t) \rangle$ can be ignored. Doing so, we find that the atomic density matrix $\rho_{\rm E}(t) = {\rm Tr}_{\rm F} ( |\psi (t) \rangle \langle \psi (t)| )$ of the emitter equals
\begin{eqnarray} \label{E14}
\rho_{\rm E}(t) &=& \left( \alpha \, |0_{\rm E} \rangle + \beta c_0(t) \, |1_{\rm E} \rangle \right)
 \left( \alpha^* \, \langle 0_{\rm E}| + \beta^* c_0(t)^* \, \langle 1_{\rm E}| \right) \notag \\
&& +  |\beta|^2 \int_{0}^{\infty} {\rm d}r \, |c_r(t)|^2  \, |0_{\rm E} \rangle \langle 0_{\rm E}| 
\end{eqnarray} 
at time $t$, where ${\rm Tr}_{\rm F} $ indicates that the trace over the states of the free radiation field is taken.   
Calculating the time derivative of $\rho_{\rm E}(t)$ with the help of the above equation, one can check that
\begin{eqnarray}
\dot \rho_{\rm E} &=& \left[ H_{\rm cond} \rho_{\rm E} - \rho_{\rm E} H_{\rm cond}^\dagger \right] 
+ \Gamma \, \sigma^-\rho_{\rm E}\sigma^+ 
\end{eqnarray}
with $H_{\rm cond}$ given in Eq.~(\ref{Hcond}). This equation is the standard quantum optical master equation of a single-photon emitter \cite{agarwal1970,petro,stokes2012}. However, notice that the above master equation has been obtained without approximations and ad hoc assumptions. The only assumption made in this paper is that the emitter resembles a point-like two-level system.

In addition, our Hamiltonian approach to quantum optical systems with photon emission reveals information about the quantum state of the emitted light. Suppose the emitter was initially excited and we only consider times $t$ that are much larger than $1/\Gamma$. In this case, the emitter and the field have already become disentangled and there is exactly one photon in the free radiation field. More concretely, the state vector $|\psi (t) \rangle $ equals $|\psi_{\rm F}(t), 0_{\rm E} \rangle$ with  
\begin{eqnarray} \label{E24newagain}
   |\psi_{\rm F}(t) \rangle = \int_0^{ct} {\rm d}r \, c_r(t) \, |r_{\rm F} \rangle 
   = \int_{- \infty}^\infty {\rm d}k \, \widetilde c_k(t) \, |k_{\rm F} \rangle \, . ~~
\end{eqnarray}
Not unlike a classical antennae connected to a finite-sized battery, the emitter transfers its energy continuously into the field until all its energy is depleted. During this process, a single-photon wave packet is generated which travels away from the ``antennae" at the speed of light. The Fourier analysis of the above quantum state of the photon reveals that the light coming from a two-level system has a Lorentzian spectrum  (cf.~Fig.~\ref{fig3}(b)) in good agreement with experiments \cite{Mollow,Mollow2,Mollow3}. 

Using the analogy of the infamous Schr\"odinger's cat \cite{Axel,Erwin} and identifying an excited and a ground state emitter with an {\em alive} and a {\em dead} cat, respectively, we find that an initially alive cat becomes slowly {\em ill} until it eventually dies (cf.~Fig.~\ref{fig1}). Our calculations show that, unless someone performs a measurement to determine whether a photon is present or not, the emitter and the field are in general in a superposition state. This is in contrast to how photon emission is usually described; most people assume that the cat is either alive or dead with the transition happening spontaneously at a random time. Indeed, there are many different ways of unravelling the dynamics generated by quantum optical master equations into individual trajectories. Which unravelling is relevant depends on the experimental circumstances. For example, in the case of continuous environment-induced measurements \cite{Moelmer,Hegerfeldt,Carmichael}, the first term in Eq.~(\ref{E14}) describes the conditional dynamics of an emitter without photon emission, while the second term can be attributed to the detection of a photon. However, this paper also demonstrates that the dynamics of the emitter are independent of the presence or absence of an observer, as it should be.

In addition to providing new insights into photon emission, our approach has immediate implications for quantum technology applications. For example, it allows us to model far-field interference effects which are essential to quantum computing schemes like the one described in Refs.~\cite{RUS,Kok}. Moreover, our analysis suggests that it is possible to apply quantum control to the state of individual emitters and to apply an antidote to an ill cat to keep it alive without the need for quantum feedback control \cite{Axel}. Our approach also allows for a stronger focus on the properties of the emitted light, including the theoretical modelling of pulse shaping of emitted photons \cite{Axel2}. More importantly for us, this paper provides novel tools for the description of more complex quantum optical systems, like atoms on opposite sides of a partially-transparent mirror surface with quantum sensing applications \cite{Dawson2021}. 

\section{Methods} \label{sec2}

\subsection{Light radiating from a point source} 

To quantise photonic wave packets originating from a point-like source, we proceed as in Refs.~\cite{photons,Jake2021,Gabriel2025} and start by noticing that they can be decomposed into so-called blips which stands for {\em bosons localised in position}. Each blip is a localised carrier of light, travels along a one-dimensional axis and has a well defined direction of propagation ${\boldsymbol s} \in {\cal S}$ and a well-defined polarisation $\lambda = {\sf H},{\sf V}$. Here ${\cal S}$ denotes the set of all possible unit vectors in three dimensions. Suppose moreover that $r \in (-\infty,\infty)$ characterises the distance of the blip from the source, with $r$ being negative and positive for light travelling towards and away from the source, respectively. Using this notation, we can characterise each blip at any given time $t$ by a set $({\boldsymbol s},\lambda, r)$ of three independent parameters. This allows us to associate each blip with an annihilation operator $a_{{\boldsymbol s}\lambda}(r)$. Since blips with different $({\boldsymbol s},\lambda, r)$ parameters are distinguishable, their annihilation operators must obey the bosonic commutator relations
\begin{eqnarray} \label{comm1}
\big[a_{{\boldsymbol s}\lambda}(r), a^\dagger_{{\boldsymbol s}'\lambda'}(r') \big] 
&=& \delta^2({\boldsymbol s}-{\boldsymbol s}') \, \delta_{\lambda \lambda'} \, \delta(r-r') \, . ~~
\end{eqnarray}
The above commutator relation ensures that the single excitation states $a^\dagger_{{\boldsymbol s}\lambda}(r) \, |0_{\rm F} \rangle$ are pairwise orthogonal and therefore distinguishable \cite{photons}. The $a_{{\boldsymbol s}\lambda}(r)$ operators can be used to represent the quantum states of all possible photonic wave packets originating from the same point-like source. Moreover, notice the inverse unit of the blip annihilation operators is distance multiplied with a solid angle segment. 

Next we have a closer look at the complex magnetic and electric field vectors $\boldsymbol{\mathcal B}_{\boldsymbol s}(r)$ and $\boldsymbol{\mathcal E}_{\boldsymbol s}(r)$ at a position $r$ in each solid angle segment. Taking the specific symmetries of light originating from a single point source into account and comparing the field observables with the observables of light propagating in one dimension, we conclude that these can be written as
\begin{eqnarray} \label{fields}
\boldsymbol{\mathcal{B}}_{\boldsymbol s}(r) &=& {1 \over c \, |r|} \sum_{\lambda = {\sf H}, {\sf V}} \int_{-\infty}^\infty {\rm d}r' \, \mathcal{R}(r,r') \, a_{{\boldsymbol s}\lambda}(r') \, {\boldsymbol s} \times {\boldsymbol {\rm e}}_{{\boldsymbol s}\lambda} \, , \notag \\
\boldsymbol{\mathcal{E}}_{\boldsymbol s}(r) &=& {1 \over |r|} \sum_{\lambda = {\sf H}, {\sf V}} \int_{-\infty}^\infty {\rm d}r' \, \mathcal{R}(r,r')  \, a_{{\boldsymbol s}\lambda}(r') \, {\boldsymbol {\rm e}}_{{\boldsymbol s}\lambda} \, .
\end{eqnarray}
The ${\boldsymbol {\rm e}}_{{\boldsymbol s} \lambda} $ in the above equation are polarisation vectors that are orthogonal to each other and to ${\boldsymbol s}$. The factor $1/|r|$ accounts for energy conservation which causes electric field amplitudes to decrease as the distance $|r|$ from the source and the surface area that they occupy increase. As we shall see below, the regularisation function $\mathcal{R}(r,r')$ in the above equations equals 
\begin{eqnarray} \label{RRR}
\mathcal{R}(r,r') &=& - \left( {\hbar c \over 4 \pi \varepsilon} \right)^{1/2} \cdot {1 \over |r-r'|^{3/2}} 
\end{eqnarray}
to ensure that each photon coming from an emitter with transition frequency $\omega_0$ has the energy $\hbar \omega_0$, i.e.~the initial energy of its source.

Before we demonstrate that the expectation values of the above field observables evolve as predicted by Maxwell's equations of classical electrodynamics, let us verify the correctness of Eq.~(\ref{RRR}). As shown in Ref.~\cite{Gabriel2025}, the energy of the electromagnetic field travelling along a given axis in one dimension can be obtained by integrating over electric and magnetic field contributions. Since energy is additive, the energy observable $H_{\rm FE}$ of light originating from a point source can be obtained by integrating over the energy contributions of light with a well defined direction ${\boldsymbol s}$. At a distance $r$ from the source, the light covers the area $r^2 \, {\rm d}^2 s$. Hence
\begin{eqnarray}
H_{\rm FE} &=& \int_{{\cal S}} {\rm d}^2{\boldsymbol s} \, H_{\rm FE} ({\boldsymbol s})
\end{eqnarray}
with the individual ${\boldsymbol s}$ contributions given by
\begin{eqnarray}
H_{\rm FE}({\boldsymbol s}) &=& \int_{-\infty}^\infty {\rm d} r \, {r^2 \over 4}  
\left[ \varepsilon \, {\boldsymbol {\cal E}}^\dagger_{\boldsymbol s}(r) \cdot {\boldsymbol {\cal E}}_{\boldsymbol s}(r) 
+ {1 \over \mu} \, {\boldsymbol {\cal B}}^\dagger_{\boldsymbol s}(r) \cdot {\boldsymbol {\cal B}}_{\boldsymbol s}(r) \right] \notag \\
\end{eqnarray}
in analogy to Eq.~(9) in Ref.~\cite{Gabriel2025}. When substituting Eq.~(\ref{fields}) into the above expression, we see that the energy observable $H_{\rm FE}({\boldsymbol s})$ is formally the same as the energy observable for light travelling along a single axis specified by ${\boldsymbol s}$. In particular,
\begin{eqnarray}
\label{energyposition}
H_{\rm FE}(\bf s) &=& \frac{\varepsilon}{2}\sum_{\lambda = {\sf H}, {\sf V}}\int_{-\infty}^{\infty}\text{d}r\int_{-\infty}^{\infty}\text{d}r'\int_{-\infty}^{\infty}\text{d}r'' \nonumber\\
&& \times \mathcal{R}(r,r')\mathcal{R}^*(r,r'')\, a^\dagger_{{\boldsymbol s}\lambda}(r')a_{{\boldsymbol s}\lambda}(r'')
\end{eqnarray}
which suggests that ${\cal R}(r,r')$ and ${\cal R}(x-x')$ in Eq.~(27) in Ref.~\cite{Gabriel2025} are the same after replacing $r$ with $x$ and $r'$ with $x'$. This is indeed the case for the regularisation function ${\cal R}(r,r')$ in Eq.~(\ref{RRR}).

To illustrate the consistency of Eq.~(\ref{RRR}) with standard quantum electrodynamics approaches more explicitly, we now calculate $H_{\rm FE}({\boldsymbol s})$ in momentum space. In momentum space, the annihilation operators $\widetilde{a}_{{\boldsymbol s} \lambda}(k)$ are the Fourier representations of the blip operators $a_{{\boldsymbol s}\lambda}(r)$,
\begin{eqnarray}
\label{FieldFT1extra}
\widetilde{a}_{{\boldsymbol s} \lambda}(k) &=& {1 \over (2\pi)^{1/2}} \int_{-\infty}^{\infty} \text{d}r \, {\rm e}^{-{\rm i}kr} \, a_{{\boldsymbol s} \lambda}(r) \, .
\end{eqnarray}
Taking into account that the regularisation distribution in Eq.~(\ref{RRR}) can also be written as 
(cf.~Fourier transform of Eq.~(37) in Ref.~\cite{Gabriel2025})
\begin{eqnarray}
\label{RegFT}
\mathcal{R}(r,r') &=& \left( \frac{\hbar c}{2 \pi^2 \varepsilon} \right)^{1/2} \int_{-\infty}^{\infty} \text{d}k \, |k|^{1/2} \, {\rm e}^{{\rm i}k(r-r')}\, , ~~
\end{eqnarray} 
and combining Eqs.~(\ref{energyposition})-(\ref{RegFT}) leads us to the energy observable
\begin{eqnarray}
\label{energymomentum}
H_{\rm FE}({\boldsymbol s}) &=& \sum_{\lambda = {\sf H}, {\sf V}}\int_{-\infty}^{\infty}\text{d}k\, \hbar c |k| \, \widetilde{a}^\dagger_{{\boldsymbol s}\lambda}(k)\widetilde{a}_{{\boldsymbol s}\lambda}(k)\, .
\end{eqnarray}
This equation demonstrates that photons with wave vector ${\boldsymbol s} |k|$ have the energy $\hbar c |k|$, as expected.

\subsection{Consistency with Maxwell's equations}

Since light in classical electrodynamics travels along straight lines, i.e.~in the respective ${\boldsymbol s}$ direction, at the speed of light $c$, we assume in the following that the same is true for the blip excitations and that 
\begin{eqnarray} \label{Solution1}
a_{{\boldsymbol s}\lambda}(r,t) = a_{{\boldsymbol s}\lambda}(r - ct,0) = a_{{\boldsymbol s}\lambda}(r - ct) 
\end{eqnarray}
in the Heisenberg picture. As we shall see below, this equation of motion guarantees that the expectation values of the electromagnetic field observables in Eq.~(\ref{fields}) evolve as predicted by Maxwell's equations. We can show that Maxwell's equations apply because the orientation of the polarisation vectors ${\boldsymbol {\rm e}}_{{\boldsymbol s} \lambda} $ with respect to the direction of propagation ${\boldsymbol s}$ has been chosen such that electric and magnetic field vectors are oriented according to the right hand rule of classical electrodynamics. In addition, we know that any wave packet travelling at the speed of light along a straight line is a solution of Maxwell's equations \cite{Jake2021,photons,Gabriel2025}. Moreover, Maxwell's equations are linear and any superposition of solutions of Maxwell's equations is therefore also a solution.

To show this more explicitly, suppose $\boldsymbol{\mathcal{E}}_{\boldsymbol s}(r,t)$ and $\boldsymbol{\mathcal{B}}_{\boldsymbol s}(r,t)$ are the observables of the complex electric and magnetic field vectors of light originating from a point-like source in the Heisenberg picture. Their expressions are the same as in Eq.~(\ref{fields}) but with the $a_{{\boldsymbol s}\lambda}(r)$ operators replaced by the $a_{{\boldsymbol s}\lambda}(r,t)$ in Eq.~(\ref{Solution1}). Given that the field vectors are always  tangential to the sphere of radius $|r|$ centred on the emitter, Maxwell's equations in spherical coordinates imply that
\begin{eqnarray}
\label{Max1}
\frac{1}{|r|}\frac{\partial}{\partial r}(|r|{\boldsymbol s} \times \boldsymbol{\mathcal{E}}_{\boldsymbol s}(r,t)) &=& -\frac{\partial \boldsymbol{\mathcal{B}}_{\boldsymbol s}(r,t)}{\partial t} \, ,\nonumber \\
\frac{c^2}{|r|}\frac{\partial}{\partial r}(|r|{\boldsymbol s} \times \boldsymbol{\mathcal{B}}_{\boldsymbol s}(r,t)) &=& \frac{\partial \boldsymbol{\mathcal{E}}_{\boldsymbol s}(r,t)}{\partial t} 
\end{eqnarray} 
where ${\boldsymbol s}$ is a constant unit vector directed away from the source. Both the electric and magnetic fields are automatically divergence-less, as they should be in free space, because ${\boldsymbol {\rm e}}_{{\boldsymbol s}\lambda}$ is orthogonal to ${\boldsymbol s}$. By substituting the field observables in Eq.~(\ref{fields}) into Eq.~(\ref{Max1}) above, one therefore finds that Maxwell's equations are satisfied when
\begin{eqnarray}
\label{Max2}
\left[\frac{\partial}{\partial r}+ \frac{1}{c}\frac{\partial}{\partial t}\right]
\int_{-\infty}^{\infty}\text{d}r' \, \mathcal{R}(r,r')\,a_{{\boldsymbol s}\lambda}(r',t) &=& 0\, .
\end{eqnarray}
By taking into account that $\mathcal{R}(r,r') = \mathcal{R}(r-r')$ due to the symmetries of the considered scenario and performing a partial integration over $r'$, we may see that this equation holds when
\begin{eqnarray}
\label{Max3}
\left[\frac{\partial}{\partial r}+ \frac{1}{c}\frac{\partial}{\partial t}\right]a_{{\boldsymbol s}\lambda}(r,t) &=& 0\, ,
\end{eqnarray}
which has the solution (\ref{Solution1}).

\subsection{The emitter-field interaction Hamiltonian}

Suppose $|0_{\rm E} \rangle$ and $|1_{\rm E} \rangle$ denote the ground and the excited states of the emitter with transition frequency $\omega_0$, respectively. Then the Hamiltonian $H_{\rm E}$ of the emitter in Eq.~(\ref{E1}) can be written as 
\begin{eqnarray}
H_{\rm E} &=& \hbar \omega_0 \, \sigma^{+}\sigma^{-} 
\end{eqnarray}
with the atomic raising and lowering operators $\sigma^{\pm}$ defined as $\sigma^{+} = |1_{\rm E} \rangle \langle 0_{\rm E}|$ and $\sigma^{-} = |0_{\rm E} \rangle \langle 1_{\rm E}|$. The only assumption that we make in the following calculations is that the dimensions of the emitter are much smaller than its transition wavelength $\lambda_0$. Demanding locality and consistency with thermodynamics \cite{stokes2017,RevMod}, the interaction Hamiltonian $H_{\rm int}$ between emitter and field can be written as 
\begin{eqnarray}  \label{InteractionHamiltonianfull}
H_{\rm int} &=& \sum_{\lambda = {\sf H}, {\sf V}} \int_{\cal S} {\rm d}^2 {\boldsymbol s} \, \hbar g_{{\boldsymbol s}\lambda} \, a_{{\boldsymbol s}\lambda}^\dagger (0) \sigma^- + {\rm H.c.}
\end{eqnarray}
with $g_{{\boldsymbol s}\lambda}$ denoting the (complex) emitter-field coupling constants. The dependence of $g_{{\boldsymbol s}\lambda}$ on ${\boldsymbol s}$ and $\lambda$ depends on the type of emitter that is being considered. Indeed, many different types of multi-polar transitions are possible \cite{multi,multi2,multi3}. For example, in the case of a dipole transition, no light is emitted in the direction of the dipole; most light escapes the emitter in the directions that are orthogonal to its dipole moment. Since the coupling constants $g_{{\boldsymbol s}\lambda} $ that we consider here can assume any value, our approach avoids standard approximations, like the usual dipole approximation. 

Having a closer look at $H_{\rm int}$, we see that a single two-level system couples effectively only to a single field annihilation operator $a(0)$. This annihilation operator is a superposition of local blip annihilation operators $a_{{\boldsymbol s}\lambda}(0)$. In the following, we therefore define annihilation operators $a(r)$ such that
\begin{eqnarray}  \label{InteractionHamiltonian2}
a(r) &=& {1 \over g} \sum_{\lambda = {\sf H}, {\sf V}} \int_{\cal S} {\rm d}^2 {\boldsymbol s} \, g_{{\boldsymbol s}\lambda} \, a_{{\boldsymbol s}\lambda}^\dagger (r) 
\end{eqnarray}
with $|g|^2 = \sum_{\lambda = {\sf H}, {\sf V}} \int_{\cal S} {\rm d}^2 s \, |g_{{\boldsymbol s}\lambda}|^2$. Using this notation, $H_{\rm int}$ in Eq.~(\ref{InteractionHamiltonian2}) simplifies to the interaction Hamiltonian in Eq.~(\ref{InteractionHamiltonian}) with $g$ representing an effective (red) emitter-field coupling constant. With the help of Eq.~(\ref{comm1}), we can check that the $a(r)$ are bosonic operators with 
\begin{eqnarray}
\big[ a(r),a^\dagger(r') \big] &=& \delta(r-r') \, . 
\end{eqnarray}
The same applies to the corresponding annihilation operators $\widetilde{a}(k)$ with
\begin{eqnarray}
	\label{FieldFT1}
	\widetilde{a}(k) &=& {1 \over (2\pi)^{1/2}} \int_{-\infty}^{\infty} \text{d}r \, {\rm e}^{-{\rm i}kr} \, a(r) 
\end{eqnarray}
of monochromatic photons in momentum space.

\subsection{The dynamics of $a(r)$ in free space}

In the absence of any emitters,  the blip excitations of the quantised electromagnetic field simply travel along straight lines at the speed of light, $c$, as shown in Eq.~(\ref{Solution1}). By comparing the dynamics of blips travelling along the $x$ axis with these dynamics \cite{photons,Jake2021,Gabriel2025}, we conclude that the field Hamiltonian $H_{\rm F}$ of light originating from a point-like source at the origin equals
\begin{eqnarray} \label{eq:dynamical_hamfull}
	H_{\rm F} = - {\rm i} \hbar c \sum_{\lambda = {\sf H}, {\sf V}} \int_{\cal S} {\rm d}^2 {\bf s} \int_{-\infty}^{\infty}\text{d}r \, a_{{\boldsymbol s}\lambda}^\dagger (r) \, {\partial \over \partial r} \,  a_{{\boldsymbol s}\lambda}(r) \, ,
\end{eqnarray}
which is formally the same as the field Hamiltonian for the one-dimensional field \cite{Gabriel2025}. In the following, we are only interested in the dynamics of the $a(r)$ operators, which allows us to write $H_{\rm F} $ as
\begin{eqnarray} \label{eq:dynamical_ham}
	H_{\rm F} &=& - {\rm i} \hbar c \int_{-\infty}^{\infty}\text{d}r \, a^\dagger (r) \, {\partial \over \partial r} \,  a(r) \, .
\end{eqnarray}
The analogy of light propagation along the $x$ axis moreover suggests that the above field Hamiltonian can be diagonalised.  Using the bosonic annihilation operators $\widetilde a(k)$ in Eq.~(\ref{FieldFT1}), $H_{\rm F}$ simplifies to the more familiar form 
\begin{eqnarray} \label{eq:dynamical_hamnew}
	H_{\rm F} &=& \int_{-\infty}^{\infty}\text{d}k \,  \hbar c k \, \widetilde a^\dagger(k) \widetilde a(k) \, .
\end{eqnarray}
This Hamiltonian has positive and negative eigenvalues and is the generator of the free-space dynamics of the photons originating from a point-like emitter. For example, $U_{\rm F}(t,0) a(r) U^\dagger_{\rm F}(t,0) = a(r + ct)$ where $U_{\rm F}(t,0)$ denotes the free-space time evolution operator. Hence $H_{\rm F}$ must be closely linked to the energy of these photons. Since energy is always positive, we assume in the following as in Ref.~\cite{Gabriel2025} that the energy observable $H_{\rm FE}$ of the photons equals $H_{\rm F}$ for positive $k$ and $-H_{\rm F}$ for negative $k$ which leads us to Eq.~(\ref{eq:dynamical_hamnewextra}).

\subsection{Dyson series expansion} \label{App:Dyson}

To simplify the following calculations, let us temporarily move into the interaction picture with respect to $t=0$ and the free Hamiltonian $H_0 = H_{\rm E} + H_{\rm F}$. In this picture, the state vector $|\psi_{\rm I} (t) \rangle$ of emitter and field equals $
|\psi_{\rm I} (t) \rangle = U_0^\dagger (t,0) \, |\psi (t) \rangle$. Here $|\psi (t) \rangle$ is the state vector in the Schr\"odinger picture and $U_0 (t,0) $ is the time evolution operator associated with $H_0$. Using the Schr\"odinger equation, we find that $|\psi_{\rm I}(t)\rangle$ also evolves according to the Schr\"odinger equation but with the time-dependent interaction Hamiltonian
\begin{eqnarray}  \label{II2}
    H_{\rm I}(t) &=& U_{0}^\dagger(t,0) \, H_{\rm int} \, U_{0}(t,0) \, .
\end{eqnarray}
The corresponding time evolution operator $U_{\rm I}(t,0)$ in the interaction picture obeys the relation 
\begin{eqnarray} \label{NotHint}
    U_{\rm I}(t,0) &=& U(0,0) + \int^t_0 {\rm d}t_1 \, \dot{U}_{\rm I}(t_1,0) \notag \\
    &=& 1 - \frac{\rm i}{\hbar} \int^{t}_{0} {\rm d}t_1 \, H_{\rm I} (t_1) U_{\rm I} (t_1,0) \, .
\end{eqnarray}
Iterating the above equation infinitely many times yields the Dyson series expansion
\begin{eqnarray} \label{stillfine}
    U_{\rm I}(t,0) &=& 1 - \frac{\rm i}{\hbar} \int^{t}_{0} {\rm d}t_1 \, H_{\rm I}(t_1) + \ldots \notag \\
    && \hspace*{-0.5cm} + \left(- {{\rm i} \over \hbar} \right)^n \int^{t}_{0} {\rm d} t_1 \ldots  \int^{t_{n-1}}_{0} {\rm d}t_n \, H_{\rm I}(t_1) \ldots H_{\rm I}(t_n) \notag \\
    && \hspace*{-0.5cm} + \ldots \, .
\end{eqnarray}
Returning into the Schr\"odinger picture, we therefore find that $U(t,0)$ can be written as
\begin{eqnarray} \label{FinalDyson}
    U(t,0) &=& \sum_{n=0}^\infty U_n (t,0) 
\end{eqnarray}
without any approximations and with the non-unitary (i.e.~conditional) time evolution operators $U_n(t,0)$ with $n \ge 1$ given by \cite{Gabriel2025}
\begin{eqnarray} \label{FinalDyson2}
    U_n(t,0) &=& \left(- \frac{\rm i}{\hbar} \right)^n \int^{t}_{0} {\rm d} t_1 \ldots \int^{t_{n-1}}_{0} {\rm d}t_n \, U_{0}(t,t_1) \notag \\
    && \times   H_{\rm int} U_{0}(t_1,t_2) \ldots H_{\rm int} \, U_{0}(t_n,0) \, . ~~~~
\end{eqnarray}

\subsection{Calculation of the coefficients $c_0(t)$ and $c_r(t)$ in Eq.~(\ref{E24})} \label{appB}

Let us first have a closer look at $c_0(t)$ which is the complex coefficient of the state vector $ |0_{\rm F} , 1_{\rm E} \rangle$. To calculate this coefficient, we first notice that 
\begin{eqnarray} \label{E27}
 U_0(t,0) \, |0_{\rm F},1_{\rm E} \rangle &=& {\rm e}^{-{\rm i} \omega_0 t} \, |0_{\rm F} , 1_{\rm E} \rangle \, .
\end{eqnarray}
since $H_0=H_{\rm E} + H_{\rm F} $. Taking this into account when calculating $U_2(t,0) \, |0_{\rm F},1_{\rm E} \rangle$, we find that 
\begin{eqnarray}
    && \hspace*{-0.7cm} U_2(t,0) \, |0_{\rm F},1_{\rm E} \rangle \notag \\
    &=& - {g^2 \over c} \int^{t}_{0} {\rm d}t_1 \int^{t_1}_{0} {\rm d} t_2 \, {\rm e}^{-{\rm i}\omega_0 (t - t_1 + t_2)} \, \delta(t_1-t_2) \, |0_{\rm F},1_{\rm E} \rangle \notag \\
\end{eqnarray}
where we have also used $\langle r_{\rm F} | r_{\rm F}' \rangle = \delta(r-r')$ which results in the Delta function $\delta(t-t')$. A local field excitation created by the emitter at a time $t_1$ can only be re-absorbed if the re-absorption occurs immediately, i.e.~at the position of the source. Performing the above time integrations yields 
\begin{eqnarray}
    U_2(t,0) \, |0_{\rm F},1_{\rm E} \rangle &=&- {g^2 t \over 2c} \, {\rm e}^{-{\rm i}\omega_0 t} \, |0_{\rm F},1_{\rm E} \rangle \, . 
\end{eqnarray}
Proceeding analogously and calculating the subsequent $U_{2m}(t,0) \, |0_{\rm F},1_{\rm E} \rangle$ terms, we find that 
\begin{eqnarray} \label{evenTimeEvolve}
    U_{2m}(t,0) \, |0_{\rm F},1_{\rm E} \rangle &=& {1 \over m!} \left( - {g^2 t \over 2c} \right)^m \, {\rm e}^{-{\rm i}\omega_0 t} \, |0_{\rm F},1_{\rm E} \rangle ~~~~
\end{eqnarray}
for all integers $m \ge 0$. Adding up the above terms for all $m$, we find that the coefficient $c_0(t)$ in Eq.~(\ref{E24}) equals
\begin{eqnarray}\label{evenExponential}
    c_0 (t) = \sum^\infty_{m=0}  {1 \over m!} \left( - {g^2 t \over 2c} \right)^m \, {\rm e}^{-{\rm i}\omega_0 t} 
\end{eqnarray}
which coincides with $c_0(t)$ in Eq.~(\ref{oma}). 

To also obtain an expression for the coefficient $c_r(t)$ introduced in Eq.~(\ref{E24}), we notice that 
\begin{eqnarray} \label{E13}
    && \hspace*{-0.6cm} U_{2m+1} (t,0) \, |0_{\rm F},1_{\rm E} \rangle \notag \\
    &=& -\frac{\rm i}{\hbar} \int^{t}_{0} {\rm d}t_1 \, U_{0}(t,t_1) H_{\rm int} \, U_{2m}(t_1,0) \, |0_{\rm F},1_{\rm E} \rangle 
\end{eqnarray}
for all integer numbers $m$ with $m \ge 0$. Combining this equation with Eqs.~(\ref{InteractionHamiltonian}), (\ref{E27}) and (\ref{evenTimeEvolve}), we therefore find that
\begin{eqnarray}
	&& \hspace*{-0.6cm} U_{2m+1} (t,0) \, |0_{\rm F},1_{\rm E} \rangle \notag \\
	&=& - {{\rm i} g \over m!} \int_{0}^{t} {\rm d}t_1 \left(- {g^2 t_1 \over 2c} \right)^{m} {\rm e}^{-{\rm i}\omega_0 t_1} \, |(c(t-t_1))_{\rm F},0_{\rm E} \rangle \, . \notag \\
\end{eqnarray}
This applies since a local field excitation created at $t_1$ travels the distance $c(t-t_1)$ away from its source within a time interval $(t_1,t)$. Next, we substitute $r = c(t-t_1)$ in the above equation to show that
\begin{eqnarray}
	&& \hspace*{-0.6cm} U_{2m+1} (t,0) \, |0_{\rm F},1_{\rm E} \rangle \notag \\
	&=& - {{\rm i}g \over m! \, c} \int_{0}^{ct} {\rm d}r \left[{g^2 \over 2c} \left( {r \over c} - t \right)\right]^m \, {\rm e}^{{\rm i}\omega_0 (r/c-t)} \, |r_{\rm F},0_{\rm E} \rangle \, . \notag \\
\end{eqnarray}
After adding up all of the above terms, we find that
\begin{eqnarray} \label{E31}
c_r(t) = -  {{\rm i} g \over c} \sum^\infty_{m=0} {1 \over m!} \left[ {g^2 \over 2c} \left( {r \over c} -t \right)\right]^m {\rm e}^{{\rm i}\omega_0(r/c-t)} 
\end{eqnarray}
for $r \in (0,ct)$. For $r>ct$, the coefficients $c_r(t)$ are zero due to the speed of light being finite. The above expression confirms Eq.~(\ref{oma}).
  
\vspace*{0.5cm}
\noindent 
{\bf Acknowledgement.} A.B. would like to thank G. C. Hegerfeldt and A. Kuhn for many interesting and stimulating discussions. T.H. and D.H. acknowledge financial support from the UK Engineering and Physical Sciences Research Council EPSRC (Grant No.~2885425 and EP/W524372/1). Moreover, H.A. acknowledges financial support from The Public Authority for Applied Education and Training in Kuwait.


\section*{References}
\begin{thebibliography}{99}
\bibitem{Zoller}
Zoller, P., Marte, M. \& Walls, D. F. 
Quantum jumps in atomic systems, 
{\em Phys. Rev.} A {\bf 35}, 198 (1987).

\bibitem{Moelmer}
Dalibard, J., Castin, Y. \& Molmer, K. 
Wave-function approach to dissipative processes in quantum optics, 
{\em Phys. Rev. Lett.} {\bf 68}, 580 (1992).

\bibitem{Hegerfeldt}
Hegerfeldt, G. C.
How to reset an atom after a photon detection: Applications to photon-counting processes, 
{\em Phys. Rev.} A {\bf 47}, 449 (1993).

\bibitem{Carmichael}
Carmichael, H.
An Open Systems Approach to Quantum Optics, 
Lecture Notes in Physics, Vol. {\bf 18} (Springer, Berlin 1993).

\bibitem{jumps4}
Nagourney, W., Sandberg, J. \& Dehmelt, H.
Shelved optical electron amplifier: observation of quantum jumps, 
{\em Phys. Rev. Lett.} {\bf 56} 2797 (1986).

\bibitem{jumps5}
Sauter, T., Neuhauser, W., Blatt, R. \& Toschek, P. E. 
Observation of quantum jumps, 
{\em Phys. Rev. Lett.} {\bf 57}, 1696 (1986).

\bibitem{jumps3}
Bergquist, J. C., Hulet, R. G., Itano, W. M. \& Wineland, D. J.
Observation of quantum jumps in a single atom, 
{\em Phys. Rev. Lett.} {\bf 57}, 1699  (1986).

\bibitem{jumps}
Dehmelt, H.
Proposed $10^{14}~\Delta \nu/\nu$ laser fluorescence spectroscopy on TI$^+$ mono-ion oscillator II, 
{\em Bull. Am. Phys. Soc.} {\bf 20}, 60 (1975).

\bibitem{jumps2}
Cook, R. J. \& Kimble, H. J.
Possibility of direct observation of quantum jumps, 
Phys. Rev. Lett. {\bf 54}, 1023 (1985).

\bibitem{jumps6}
Itano, W. M., Bergquist, J. C. \& Wineland,  D. J.
Early observations of macroscopic quantum jumps in single atoms,
{\em Int. J. Mass Spectrom.} {\bf 377}, 403 (2015).

\bibitem{Beige}
Beige A. \& Hegerfeldt, G. C.
Quantum Zeno effect and light-dark periods for a single atom, 
{\em J. Phys.} A {\bf 30}, 1323 (1997). 

\bibitem{Ballentine2}
Ballentine, L. E. 
Comment on ``Quantum Zeno effect",
{\em Phys. Rev.} A {\bf 43}, 5165 (1991). 

\bibitem{Ballentine3}
W.M. Itano, D.J. Heinzen, J.J. Bollinger, and D.J. Wineland, 
Reply to ``Comment on `Quantum Zeno effect’’’
{\em Phys. Rev.} A {\bf 43}, 5168 (1991).

\bibitem{Scully}
Scully, M. O. \& Dr{\"u}hl, K. 
Quantum eraser: A pro-posed photon correlation experiment concerning observation and ”delayed choice” in quantum mechanics, {\em Phys. Rev.} A {\bf 25}, 2208 (1982). 

\bibitem{Eichmann}
Eichmann., U., Bergquist,  J. C., Bollinger, J. J., Gilligan, J. M., Itano, W. M., Wineland, D. J. \& Raizen, M. G.
Young’s interference experiment with light scattered from two atoms, 
{\em Phys. Rev. Lett.} {\bf 70}, 2359 (1993).

\bibitem{Schoen}
Sch{\"o}n, C. \& Beige, A.
Analysis of a two-atom double-slit experiment based on environment-induced measurements,
{\em Phys. Rev.} A {\bf 64}, 023806 (2001).

\bibitem{Schoen2}
Beige, A., Sch{\"o}n, C. \& Pachos, J.
Interference of spontaneously emitted photons, 
{\em Fortschr. Phys.} {\bf 50}, 594 (2002).

\bibitem{Dawson2021}
Furtak-Wells, N., Dawson, B., Mann, T., Jose, G. \& Beige, A. 
Mirror-mediated ultralong-range dipole-dipole interactions, 
{\em Opt. Quantum Electron.} {\bf 56}, 1287 (2024).

\bibitem{agarwal1970}
Agarwal, G.S. 
Quantum statistical theories of spontaneous emission and their relation to other approaches,
Springer Tracts in Modern Physics, Vol. {\bf 70} (Springer, Berlin, Heidelberg 1974).

\bibitem{petro}
Breuer H.-P. \& Petruccione, F.
The theory of open quantum systems 
(Oxford University Press, Oxford 2002).

\bibitem{stokes2012}
Stokes, A., Kurcz, A., Spiller, T. P. \& Beige, A., 
Extending the validity range of quantum optical master equations, 
{\em Phys. Rev.} A \textbf{85}, 053805 (2012).

\bibitem{Axel}
Alvarez, J.-R., IJspeert, M., Barter, O., Yuen, B., Barrett, T. D., Stuart, D., Dilley, J., Holleczek, A. \& Kuhn, A. 
How to administer an antidote to Schr{\"o}dinger’s cat,
{\em J. Phys.} B {\bf 55}, 054001 (2022).

\bibitem{Erwin}
Schr{\"o}dinger, E. 
Die gegenw{\"a}rtige Situation in der Quantenmechanik, 
{\em Sci. Nat.} {\bf 23}, 807 (1935).

\bibitem{Pohl}
M{\"u}hlschlegel, P., Eisler, H.-J., Martin, O. J. F., Hecht B. \& Pohl, D. W. 
Resonant Optical Antennas,
{\em Science} {\bf 308}, 1607 (2005).

\bibitem{RUS}
Lim, Y. L., Beige A. \& Kwek, L. C. 
Repeat-until-success linear optics distributed quantum computing,
{\em Phys. Rev. Lett.} {\bf 95}, 030505 (2005). 

\bibitem{Kok}
Barrett S. D. \& Kok, P.
Efficient high-fidelity quantum computation using matter qubits and linear optics,
{\em Phys. Rev.} A {\bf 71}, 060310 (2005).

\bibitem{Eberly}
Fedorov, M. V., Efremov, M. A., Kazakov, A. E., Chan, K. W., Law, C. K. \& Eberly, J. H.
Spontaneous emission of a photon: Wave-packet structures and atom-photon entanglement,
{\em Phys. Rev.} A {\bf 72}, 032110 (2005).

\bibitem{Longhi}
Longhi, S.
Virtual atom-photon bound states and spontaneous emission control,
{\em Opt. Lett.} {\bf 50}, 3026 (2025).

\bibitem{photons}
Hodgson, D., Southall, J., Purdy, R. \& Beige, A. 
Local Photons,
{\em Front. Photon.} \textbf{3}, 978855 (2022).

\bibitem{Jake2021}
Southall, J., Hodgson, D., Purdy, R. \& Beige, A. 
Locally acting mirror Hamiltonians, 
{\em J. Mod. Opt.} \textbf{68}, 647 (2021).

\bibitem{Gabriel2025}
Waite, G., Hodgson, D., Lang, B., Alapatt, V. \& Beige, A. 
Local-photon model of the momentum of light, 
{\em Phys. Rev.} A \textbf{111}, 023703 (2025).

\bibitem{Sipe}
Sipe, J. E. 
Photon wave functions,
{\em Phys. Rev.} A {\bf 52}, 1875  (1995).

\bibitem{Rempe}
Kuhn, A., Hennrich, M. \& Rempe, G.
Deterministic Single-Photon Source for Distributed Quantum Networking,
{\em Phys. Rev. Lett.} {\bf 89}, 067901 (2002).

\bibitem{multi}
Craig, D. P. \& Thirunamachandran, T.
Molecular Quantum Electrodynamics 
(Academic, London 1984).

\bibitem{multi2}
Cohen-Tannoudji, C., Dupont-Roc, J. \& Grynberg, G.
Photons and Atoms: Introduction to Quantum Electrodynamics
(Wiley-Interscience, New York 1997).

\bibitem{multi3}
Woolley, R. G.
Charged particles, gauge invariance, and molecular electrodynamics,
{\em Int. J. Quantum Chem.} {\bf 74}, 531 (1999).

\bibitem{Heg2}
Hegerfeldt, G. C.
Causality problems for Fermi’s two-atom system, 
{\em Phys. Rev. Lett.} {\bf 72}, 596 (1994).

\bibitem{Milonni}
Milonni, P. W., James, D. F. V. \& Fearn, H. 
Photodetection and causality in quantum optics,
{\em Phys. Rev.} A {\bf 52}, 1525 (1995).

\bibitem{stokes2017}
Stokes, A., Deb, P. \& Beige, A., 
Using thermodynamics to identify quantum subsystems, 
{\em J. Mod. Opt.} {\bf 64}, S7 (2017).

\bibitem{RevMod}
Stokes, A. \& Nazir, A.
Implications of gauge freedom for nonrelativistic quantum electrodynamics,
{\em Rev. Mod. Phys.} {\bf 94}, 045003 (2022).

\bibitem{Kurcz}
Kurcz, A., Capolupo, A., Beige, A., Del Giudice, E. \& Vitiello, G.
Energy concentration in composite quantum systems,
{\em Phys. Rev.} A {\bf 81}, 063821 (2010).

\bibitem{WW}
Weisskopf, V.  and Wigner, E.
Berechnung der nat{\"u}rlichen Linienbreite auf Grund der Diracschen Lichttheorie,
{\em Z. Phys.} {\bf 63}, 54 (1930).

\bibitem{Stenholm}
Stenholm, S. \& Suominen, K.-A.
Weisskopf-Wigner decay of excited oscillator states,
{\em Opt. Express} {\bf 2}, 378 (1998).

\bibitem{Berman}
Berman, P. R. \& Ford, G. W.
Spectrum in spontaneous emission: Beyond the Weisskopf-Wigner approximation,
{\em Phys. Rev.} A {\bf 82}, 023818 (2010).

\bibitem{Ornigotti}
Ornigotti, M., Conti, C. \& Szameit, A.  
Quantum X waves with orbital angular momentum in nonlinear dispersive media,
{\em J. Opt.} {\bf 20}, 065201 (2018).

\bibitem{Mollow}
Mollow, B. R.
Power Spectrum of Light Scattered by Two-Level Systems,
{\em Phys. Rev.} {\bf 188}, 1969 (1969).

\bibitem{Mollow2}
Cohen-Tannoudji, C., Diu, B. \& Lal{\"o}e, F.  
Quantum Mechanics 
(John Wiley and Sons, New York 1992)

\bibitem{Mollow3}
Scully, M. O. \& Zubairy, M. S. 
Quantum Optics
(Cambridge University Press, Cambridge 1997).

\bibitem{Abeer2}
Al Ghamdi, A. Jose, J. \& Beige, A.
Cavity QED beyond the Jaynes-Cummings model,
arXiv:2510.16634 (2025). 

\bibitem{Ding}
Ding, L., Fan, J. \& Qiu, X.
Universally Robust Control of Open Quantum Systems,
arXive:2508.0737 (2025).

\bibitem{Axel2}
Nisbet-Jones, P. B. R., Dilley, J., Ljunggren, D. \& Kuhn, A.
Highly efficient source for indistinguishable single photons of controlled shape,
{\em New J. Phys.} {\bf 13}, 103036 (2011).

\bibitem{Ballentine}
Ballentine, L. E. 
Limitations of the projection postulate,
{\em Found. Phys.} {\bf 20}, 1329 (1990).

\end{thebibliography}
\end{document}